\documentclass[a4paper]{article}
\usepackage{Odyssey2024}
\usepackage{epsfig,amssymb,amsmath, booktabs, tipa, multirow}
\ninept

\title{Speaker Embeddings With Weakly Supervised Voice Activity Detection For Efficient Speaker Diarization}

\name{Jenthe Thienpondt, Kris Demuynck}

\address{IDLab, Department of Electronics and Information Systems  \\
Ghent University - imec, Belgium \\
{\small \tt jenthe.thienpondt@ugent.be} }
\begin{document}
\maketitle

\begin{abstract}

Current speaker diarization systems rely on an external voice activity detection model prior to speaker embedding extraction on the detected speech segments. In this paper, we establish that the attention system of a speaker embedding extractor acts as a weakly supervised internal VAD model and performs equally or better than comparable supervised VAD systems. Subsequently, speaker diarization can be performed efficiently by extracting the VAD logits and corresponding speaker embedding simultaneously, alleviating the need and computational overhead of an external VAD model. We provide an extensive analysis of the behavior of the frame-level attention system in current speaker verification models and propose a novel speaker diarization pipeline using ECAPA2 speaker embeddings for both VAD and embedding extraction. The proposed strategy gains state-of-the-art performance on the AMI, VoxConverse and DIHARD III diarization benchmarks.

\end{abstract}

\section{Introduction}
Speaker diarization tries to solve the question of 'Who spoke when?' of an audio file. Current state-of-the-art speaker diarization systems~\cite{voxsrc23_dia_1st, voxsrc23_dia_2nd} follow a cascaded approach by dividing the problem statement in several subtasks, usually consisting of voice activity detection (VAD), speaker embedding extraction and embedding clustering. While recent end-to-end diarization models provide a compelling and promising alternative~\cite{eend_diarization, eend_sim_mix}, current end-to-end approaches only prove competitive with cascaded systems in a limited set of scenarios~\cite{wide_band_eend}.



Most VAD models rely heavily on artificially generated training data because of the limited availability of audio with frame-level speech/non-speech labels. For example, the VAD system provided in the popular SpeechBrain~\cite{speechbrain} library consists of a hybrid recurrent and convolutional neural network trained on a mixture of LibriSpeech~\cite{libri} and the QUT-NOISE~\cite{qut_noise} augmentation dataset. In a similar fashion, the winners of the diarization track of the recent VoxCeleb Speaker Recognition Challenge~(VoxSRC) relied on VoxCeleb1\&2~\cite{vox1, vox2} mixed with MUSAN~\cite{musan} augmentations to train their VAD model~\cite{voxsrc23_dia_1st}. While attempts are made by training VAD models directly on a combination of the available labeled diarization datasets~\cite{vad_pyannote_2.1, vad_pyannote_3.1}, performance degrades significantly on out-of-domain test data~\cite{vad_pyannote_3.1}.




In contrast to VAD models, speaker embedding extractors have access to a large number of labeled speech corpora, such as VoxCeleb1\&2. Recent advances in deep learning architectures are capable of exploiting this abundance of data, as shown by the popular x-vector~\cite{x_vectors} and ECAPA-TDNN~\cite{ecapa_tdnn} architectures based on time-delayed neural networks~(TDNNs) or speech-adapted variations of ResNet models~\cite{freq_paper, magneto}. Recently, hybrid models combining both architectures, such as CNN-TDNN~\cite{NetEase_CNN_TDNN} and ECAPA2~\cite{ecapa2}, have shown additional performance benefits.







In this paper, we propose a novel, weakly supervised approach for VAD which exploits the notion that the frame-level attention mechanism of a speaker embedding extractor shares the goal of a VAD model, i.e. indicating intelligible speech frames. An attention mechanism allows speaker embedding  extractors to weight individual frame-level features before the temporal pooling operation, highlighting frames containing speaker information while suppressing non-speech or noisy frames. By interpreting the attention weights as frame-level VAD logits, the speaker embedding extractor can act as a weakly supervised VAD model, only relying on utterance-level speaker labels during training. The potential benefits of this strategy are substantial compared to a traditional supervised VAD model. It alleviates the need of an (artificial) dataset with frame-level speech/non-speech labels, can learn from the large availability of speaker verification corpora, reduces the computational impact of the VAD step to zero when also used as embedding extractor for diarization and requires no or minimal modifications to current speaker verification architectures and training setups. 


The rest of the paper is organized as follows: Section 2 describes the derivation of VAD logits from a frame-level attention mechanism in speaker embedding extractors. Section 3 analyzes these attention-based VAD logits from a pre-trained ECAPA2 model to asses robustness, resolution and behavior in a variety of speech scenarios. Section 4 uses these insights to propose a diarization pipeline using a single ECAPA2 model for both VAD and embedding extraction. Subsequently, Section 5 explains the experimental setup for testing the VAD, speaker embedding and final diarization performance separately with the corresponding results given in Section 6. Finally, Section 7 provides some concluding remarks.


\begin{figure*}[t]
\includegraphics[width=1.0\textwidth]{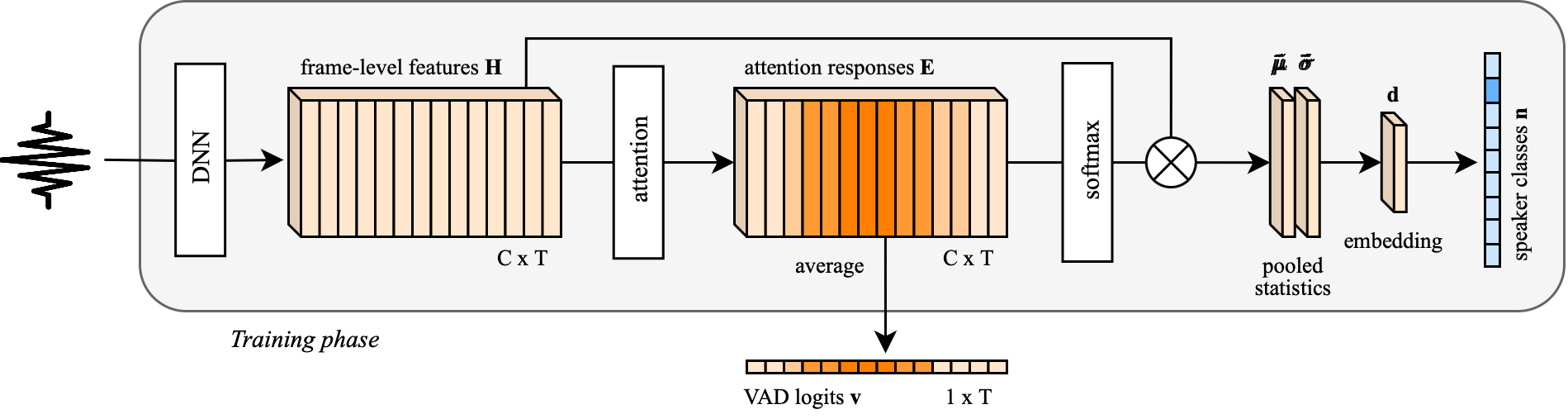}
\caption{{\it Temporal- and channel-based attention mechanism of a speaker embedding extractor during training. The attention responses $\pmb{E}$ can be interpreted as VAD logits, resulting in a weakly supervised VAD training setup only requiring utterance-level speaker labels.}}
\label{fig:vad_training}
\end{figure*}

\section{Weakly supervised VAD}
\label{sec:weakly_supervised_vad}

Current state-of-the-art speaker embedding extractors often employ a frame-level attention mechanism before the temporal pooling layer~\cite{ecapa_tdnn}. The usage of such attention systems is commonly motivated by its ability to highlight frame-level features containing speaker information and suppressing frames with less relevant information such as silence, noise or distorted speech. In this section, we start with a brief overview of the temporal- and channel-based attention mechanism used in state-of-the-art speaker embedding extractors such as ECAPA2 and how the attention responses can be transformed to VAD logits. 




\subsection{Temporal- and channel-based attention}

The simplest form of attentive statistical pooling in speaker verification architectures produces a singular attention scalar $e_{t}$ for each frame-level feature $\pmb{h}_{t}$ at time step $t$ in the statistical pooling layer. This attention model can be extended to the channel dimension, as proposed in~\cite{ecapa_tdnn}, by calculating a separate scalar $e_{t, c}$ for each frame-level feature channel $c$:

\begin{equation}
\label{channel_stat_pool}
e_{t, c} = \pmb{p}^{T}_{c} f(\pmb{W}\pmb{h}_{t} + \pmb{b}) + k_{c}
\end{equation}

with $\pmb{W} \in \mathbb{R}^{R \times C}$ and $\pmb{b} \in \mathbb{R}^{R \times 1}$ being the respective weights and bias of an intermediate linear bottleneck layer of dimension $R$ followed by the ReLU non-linearity given by $f(.)$. Subsequently, a linear operation with weights $\pmb{p} \in \mathbb{R}^{R \times 1}$ and bias $k$ produces the attention scalar $e_{t, c}$ which is normalized across the temporal dimension by applying a softmax function:

\begin{equation}
\label{stat_pool_softmax}
\alpha_{t,c} = \frac{\exp(e_{t,c})}{\sum_{t}^{T} \exp(e_{t,c})}.
\end{equation}

with $\alpha_{t, c}$ being the normalized attention scalar used to weight the frame- and channel-level feature $h_{t, c}$. The attentive mean pooled statistics $\tilde{\mu}_{c}$ for each channel $c$ can now be calculated as follows:

\begin{equation}
\label{stat_pool_mu}
\tilde{\mu}_{c} = \sum_{t}^{T}\alpha_{t,c}h_{t,c}.
\end{equation}

Correspondingly, the attentive standard deviation $\tilde{\sigma}_{c}$ is given by:

\begin{equation}
\label{stat_pool_mu_sigma}
\tilde{\sigma}_{c} = \sqrt{\sum_{t}^{T} \alpha_{t,c}h_{t,c}^{2}-\tilde{\mu}_{c}^{2}}.
\end{equation}

The concatenation of $\pmb{\tilde{\mu}}$ and $\pmb{\tilde{\sigma}}$ can subsequently be projected to a lower dimension using a linear layer to produce the final speaker embedding $\pmb{d}$. Figure~\ref{fig:vad_training} depicts the temporal- and channel-based attention mechanism during the training phase of a speaker embedding extractor.

\begin{figure}[t]
\centering
\includegraphics[width=\columnwidth]{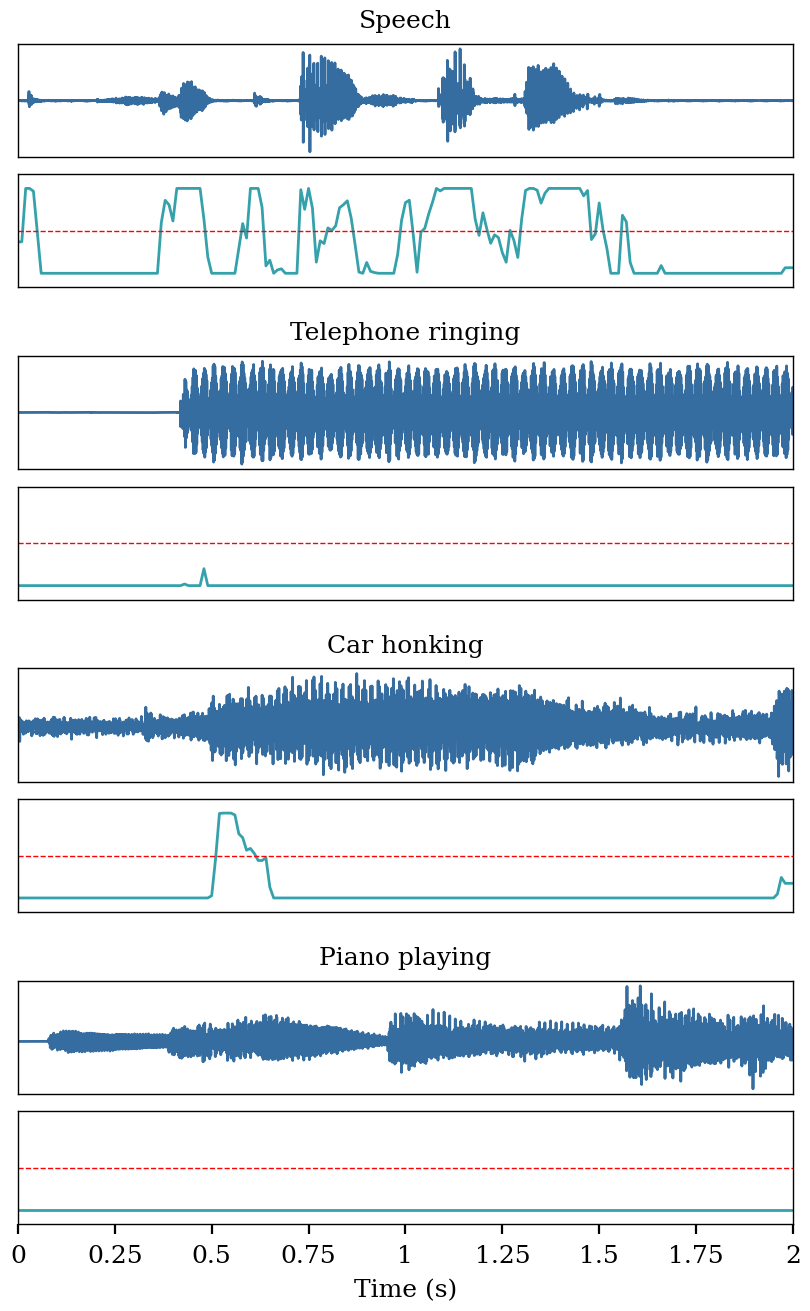}
\caption{{\it ECAPA2 mean frame-level attention responses $\pmb{v}$ (green) on speech and non-speech audio inputs (blue). The $\pmb{v}$ values are capped between -0.3 and 0.3 for clarity. The red dotted line represents the optimal speech decision threshold. Notice that the attention system behaves similarly to a VAD model, without access to speech/non-speech labels during training.}}
\label{fig:attention_weights}
\end{figure}

\begin{figure}[t]
\includegraphics[width=\columnwidth]{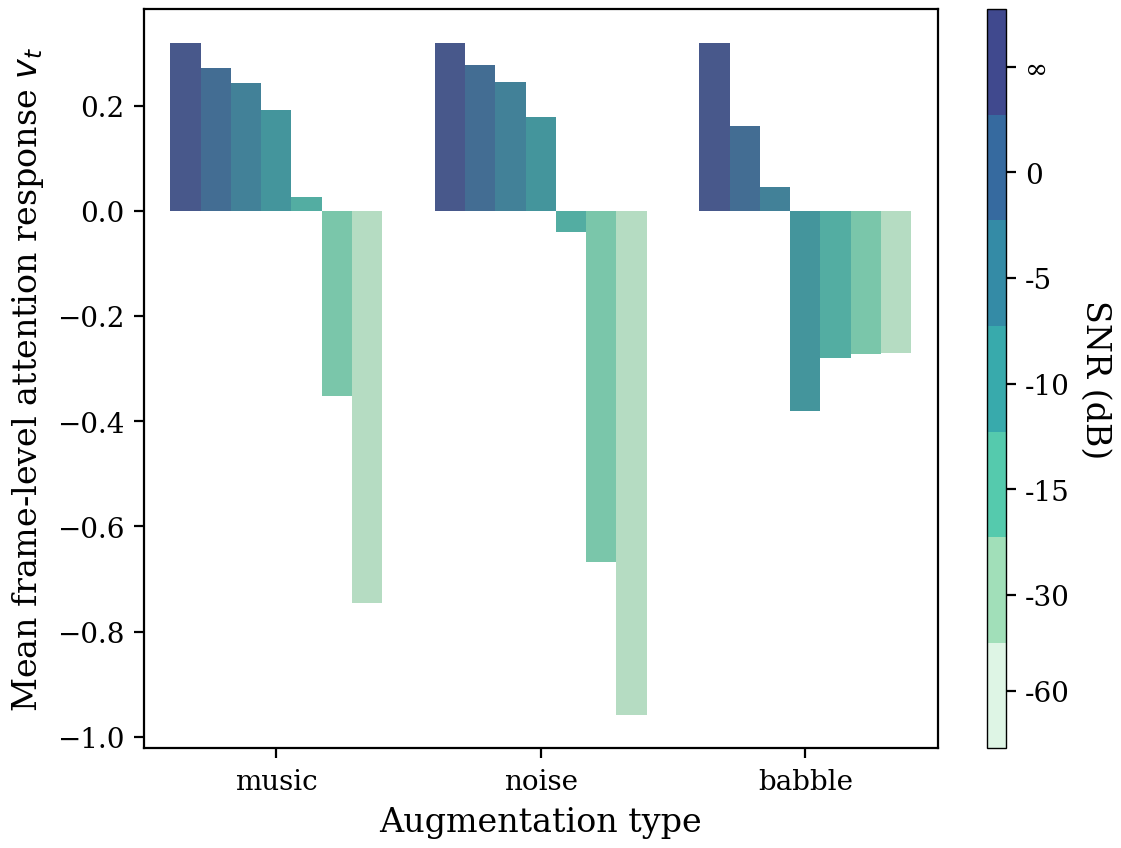}
\caption{{\it Mean attention response $v_t$ from LibriSpeech samples augmented with noise, music and babble at different SNR ratios.}}
\label{fig:attention_snr}
\end{figure}

\begin{figure}[t]
\includegraphics[width=\columnwidth]{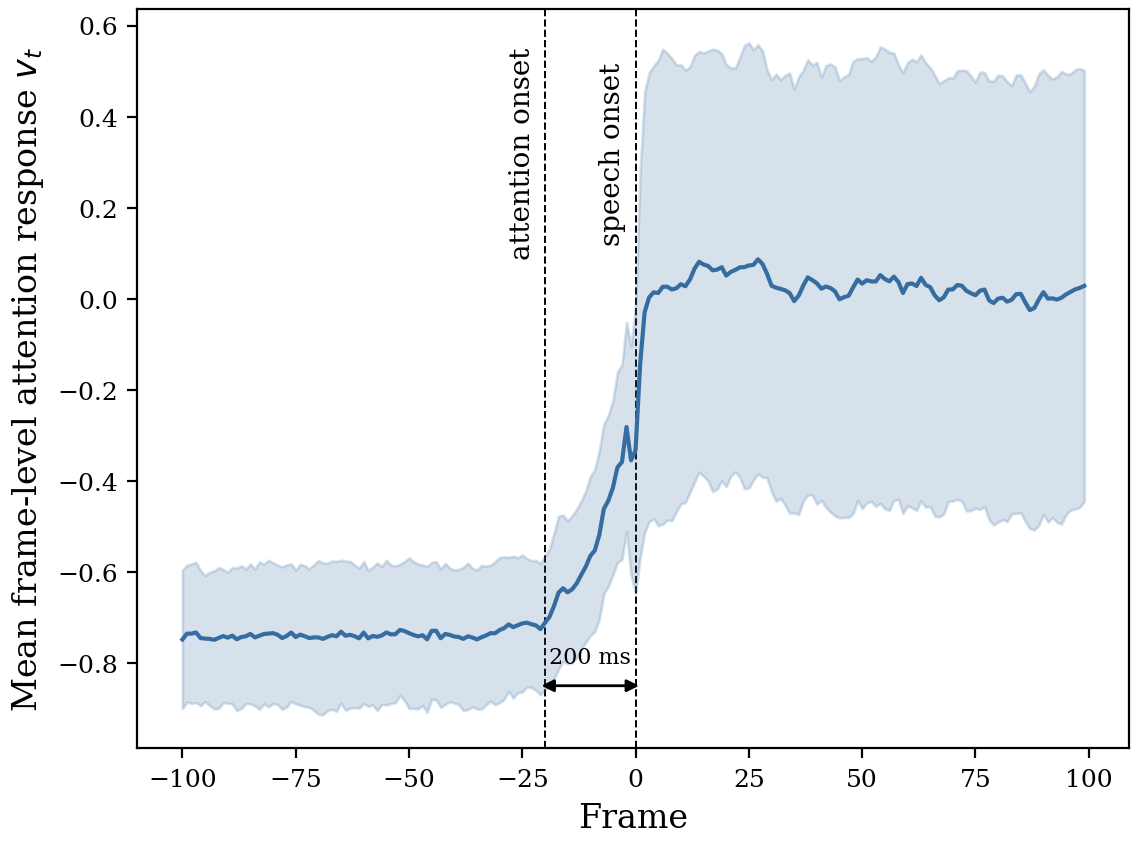}
\caption{{\it Mean attention response $v_t$ and standard deviation of frame-level features when transitioning from noise to speech.}}
\label{fig:rf_graph}
\end{figure}

\subsection{From attention to voice activity detection}
The attentive scalars in $\pmb{e}_t$ represent the importance of the corresponding frame-level feature located at $t$ to produce a robust speaker embedding. Consequently, the attention system should act similarly to a VAD model, assigning lower weights to silent, noisy and irrelevant frames while assigning higher weights to frames with distinguishable speech. Additionally, state-of-the-art speaker embeddings are robust against a wide variety of background conditions, such as music, noise and babble interference, which should enable the model to distinguish between frames with speech in challenging conditions and true non-speech frames. To adapt the frame-level attention vector $\pmb{e}_t$ for the VAD task, we calculate the average value across the channel dimension, resulting in one VAD logit $v_t$ for each time step:

\begin{equation}
\label{vad_value_simple}
v_{t} = \frac{\sum_{c}^{C}e_{t,c}}{C}.
\end{equation}

The subsequent training setup results in a weakly supervised VAD model only depending on utterance-level speaker labels, which is a far less strict requirement as opposed to the need of frame-level speech/non-speech labels for supervised VAD training. Furthermore, the VAD logits $\pmb{v}$ can be extracted simultaneously with the speaker embedding $\pmb{d}$, at no additional computational cost. This can especially be beneficial when employed in a speaker diarization system, which will be explored in Section~\ref{sec:speaker_diarization}.


\section{Attention response analysis}


In this section, we analyze the attention responses of a speaker embedding extractor to assess robustness, resolution and behavior on a variety of speech conditions to asses the feasibility of a weakly supervised VAD model in practical scenarios. For all analyses in this section, we use a pre-trained ECAPA2 speaker verification model as proposed in~\cite{ecapa2} which employs the attention mechanism described in Section~\ref{sec:weakly_supervised_vad}. The model is trained using the development partition of VoxCeleb2, which provides audio from a wide range of background conditions and should result in a robust attention mechanism. More details about this model can be found in the accompanying paper~\cite{ecapa2}.

\subsection{Speech and non-speech responses}
Figure~\ref{fig:attention_weights} depicts the mean frame-level attention responses $\pmb{v}$ provided by the ECAPA2 model given a variety of input audio sources. We observe that the frame-level attention responses behave similarly to a VAD model, with higher values assigned to speech frames and lower values to intermediate silence frames in the speech example. Noticeably, the attention system is able to distinguish speech from non-speech audio sources, consistently assigning low weights to frames with common background augmentations. However, we could also observe some errors. The attention weights have a tendency to assign higher values to frames at the onset of an audio source, even for non-speech sounds, as depicted in the car honking example. This leads to some occasional classification errors of a small set of frames of non-speech sources.

\subsection{Augmentation robustness}
To assess the impact of common background audio on the ECAPA2 attention mechanism, we depict the mean $v_t$ value of speech frames from 1000 random LibriSpeech~\cite{libri} utterances augmented with music, babble and noise audio at varying signal-to-noise~(SNR) levels. Only frames with speech are included to prevent silent frames impacting the mean $v_t$.

We observe that noise and music augmentation behaves similarly, with a drastic decrease in the detection of speech frames when the SNR drops below -30, at which point the speaker becomes barely audible. Babble proves more challenging, as the presence of other speakers quickly results in a lower $v_t$ response. When interfering speakers become sufficiently prominent, the embedding extractor will have increased difficulty to determine a dominant speaker, ultimately assigning a low weight to the frame. As a result, the $v_t$ response will likely be less useful to determine overlapping speech frames.

\begin{figure*}[t]
\includegraphics[width=1.0\textwidth]{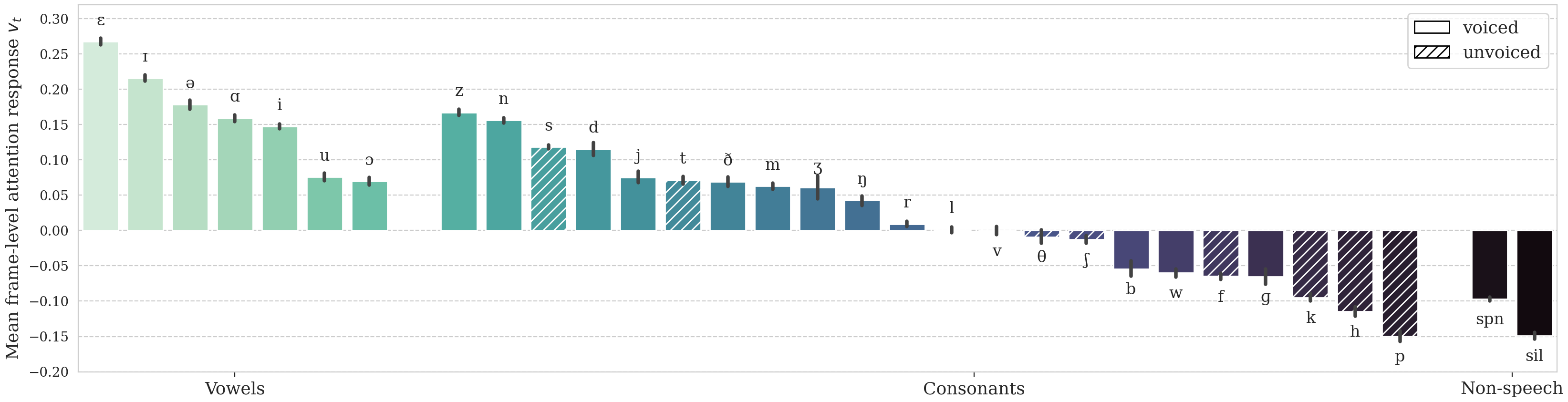}
\caption{{\it Attention responses $v_t$ on TIMIT dataset, grouped by phoneme. 'spn' and 'sil' indicate speaker noise and silence, respectively.}}
\label{fig:phone_response}
\end{figure*}

\subsection{Attentive region of influence}
Frame-level features before the pooling operation in current state-of-the-art speaker embedding extractors have a wide receptive field due to a large number of stacked convolutional operations, amplified by striding or dilation. In comparison, typical VAD models have a more narrow receptive field due to the goal of providing frame-level speech probabilities, which can potentially be negatively impacted if the decision is based on a too wide input region. However, it is known that frame-level features of speaker embedding extractors are much more inclined to be influenced by the corresponding input features at the center of their receptive field~\cite{ecapa2}.

We analyze the region of influence for the $v_{t}$ response in Figure~\ref{fig:rf_graph}, which depicts the average $v_{t}$ value and corresponding standard deviation of frames when transitioning from silence to speech of 1000 random LibriSpeech utterances. We observe empirically that the region of influence of the $v_{t}$ value is on average $[t-20, t+20]$, with each frame corresponding to 10 ms of audio, and more heavily weights central input frames. This is significantly smaller than the corresponding receptive field of $v_{t}$ and proves favorable for the combination of fine-grained VAD and robust utterance-level speaker embeddings.

\subsection{Phoneme response}
A VAD model should be able to react similarly to vowels and consonants, although the latter is more challenging due to the transient and abrupt nature of consonants compared to vowels. To gain a better understanding on how the attention mechanism of a speaker embedding extractor reacts to different acoustic units, Figure~\ref{fig:phone_response} illustrates the average $v_{t}$ response, grouped by phoneme, of audio from the training partition of TIMIT~\cite{timit}.

\begin{figure}[t!]
\includegraphics[width=\columnwidth]{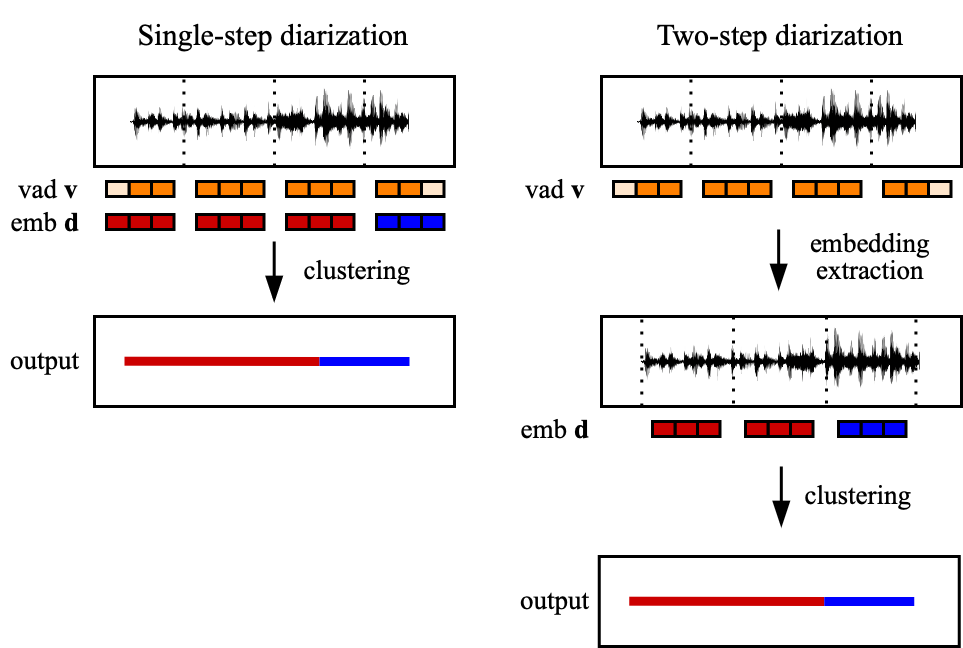}
\caption{{\it Single-step diarization with combined embedding and VAD extraction (left) and standard two-step diarization (right).}}
\label{fig:window_alignment}
\end{figure}

We can observe that the attention system is significantly more inclined to put higher $v_t$ values to frames containing vowels in comparison to consonants. We suspect that vowels tend to carry more speaker-specific information. For example, the more defined formants of vowels should provide a better fingerprint of the vocal tract and corresponding speaker identity. Furthermore, there is a clear distinction between voiced and unvoiced consonants, with some unvoiced consonants given a similar speaker-identifying weight as silence (e.g. \textipa{p}). However, co-articulation of unvoiced consonants with other phonemes possibly mitigates the overall impact on practical VAD performance.



\section{Single-step speaker diarization}
\label{sec:speaker_diarization}



As established in the previous section, the attention system of a speaker embedding extractor behaves similarly to a VAD model. This opens up the possibility to use a single speaker verification model for both VAD and embedding extraction, removing the computational overhead of the VAD phase. Subsequently, we propose a full diarization pipeline using a single speaker verification model for both VAD and embedding extraction, which we will refer to as single-step diarization.

Standard cascaded diarization performs frame-level VAD and speaker embedding extraction in separate steps, with the windowed embedding extraction done on the detected speech segments. To enable single-step VAD and embedding extraction, we obtain both the frame-level $v_{t}$ values and corresponding speaker embedding using a sliding window of width $w$ and step size $s$ over the input utterance. Subsequently, we determine the speech segments by averaging the overlapping $v_{t}$ values and performing hysteresis thresholding with starting and stopping thresholds $\theta_{on}$ and $\theta_{off}$, respectively. Afterwards, speaker embeddings from windows with no speech segments detected are discarded and not used in the clustering phase. When speaker labels are provided by the clustering algorithm, only the corresponding detected speech frames gets assigned to this speaker. Figure~\ref{fig:window_alignment} depicts the difference between our single-step and standard two-step VAD and embedding extraction.

\begin{table*}
  \caption{{\it VAD error rate on various test sets using the attention responses of ECAPA2 in comparison with other publicly available supervised (SV) VAD models. \pmb{Bold} indicates best VAD performance of models not using in-domain training data.}}
  \label{tab:vad_results}
  \vspace{2mm}
  \centering
  \begin{tabular}{lccccc|ccc|ccc|ccc}
    \toprule
    \multirow{2}{*}{\textbf{System}} &
    \multirow{2}{*}{\textbf{SV}} &
    \multirow{2}{*}{\textbf{In-domain}} &
    \multicolumn{3}{c}{\textbf{AMI (array)}} &
    \multicolumn{3}{c}{\textbf{AMI (headset)}} &
    \multicolumn{3}{c}{\textbf{VoxConverse}} &
    \multicolumn{3}{c}{\textbf{DIHARD III}} \\
    \cmidrule(lr){4-6} \cmidrule(lr){7-9} \cmidrule(lr){10-12} \cmidrule(lr){13-15}
    \multicolumn{1}{c}{\textbf{}} &
    \multicolumn{1}{c}{\textbf{}} & 
    \multicolumn{1}{c}{\textbf{}} & 
    \multicolumn{1}{c}{\textbf{FA}} & \multicolumn{1}{c}{\textbf{MS}} & \multicolumn{1}{c}{\textbf{VAD}} &
    \multicolumn{1}{c}{\textbf{FA}} & \multicolumn{1}{c}{\textbf{MS}} & \multicolumn{1}{c}{\textbf{VAD}} &
    \multicolumn{1}{c}{\textbf{FA}} & \multicolumn{1}{c}{\textbf{MS}} & \multicolumn{1}{c}{\textbf{VAD}} &
    \multicolumn{1}{c}{\textbf{FA}} & \multicolumn{1}{c}{\textbf{MS}} & \multicolumn{1}{c}{\textbf{VAD}} \\
    \midrule
    SpeechBrain & \checkmark & - & 4.3 & 1.4 & 5.7 &  3.5 & 1.5 & 5.0  & 2.2 & 1.7 & 3.9 & 17.8 & 3.6 & 21.4 \\
    Silero & \checkmark & - & 3.6 & 2.3 & 5.9 & 3.3 & 1.5 & 5.2 & 2.4 & 1.9 & 4.3 & 11.0 & 9.1 & 20.1 \\
    PyAnnote 3.1 & \checkmark & \checkmark & \textit{1.9} & \textit{2.4} & \textit{4.3} & \textit{2.0} & \textit{1.1} & \textit{3.1} & \textit{1.6} & \textit{0.6} & \textit{2.2} & \textit{3.9} & \textit{3.2} & \textit{7.1} \\
    \midrule
    ECAPA2 (VAD only) & - & - & 3.0 & 2.6 & \textbf{5.6} & 3.0 & 1.9 & \textbf{4.9} & 1.2 & 1.5 & \textbf{2.7} & 8.3 & 9.9 & \textbf{18.2} \\
    \bottomrule
  \end{tabular}
\end{table*}

\begin{table*}
  \caption{{\it Diarization error rate on various test sets using ECAPA2 in comparison to other published non-fusion and single-scale diarization results using oracle VAD. Results are given with and without usage of the oracle number of speakers.}}
  \label{tab:dia_results}
  \vspace{2mm}
  \centering
  \begin{tabular}{lcccccccccc}
    \toprule
    \multirow{2}{*}{\textbf{System}} &
    \multicolumn{2}{c}{\textbf{AMI (array)}} &
    \multicolumn{2}{c}{\textbf{AMI (lapel)}} & 
    \multicolumn{2}{c}{\textbf{AMI (headset)}} &
    \multicolumn{2}{c}{\textbf{VoxConverse}} &
    \multicolumn{2}{c}{\textbf{DIHARD III}}\\
     & oracle & estimate & oracle  & estimate & oracle & estimate & oracle & estimate & oracle & estimate \\
    \midrule
    xvec+ClusterGAN~\cite{cluster_gan} & 3.6 & 2.8 & -  & - & - & - & - & - & - & - \\
    ECAPA-TDNN~\cite{ecapa_tdnn_diarization} & 2.8 & 3.0 & 2.3  & 2.5 & 1.7 & 4.0 & - & - & - & - \\
    TitaNet-L~\cite{titanet} & - & - & 2.0 & 2.0 & 1.7 & 1.9 & - & - & - & -\\
    DR-DESA~\cite{dim_reduction_diarization} & - & - & - & - & - & - & - & 4.4 & - & 15.0 \\
    \midrule
    ECAPA2 & \textbf{2.0} & \textbf{1.8} & \textbf{1.8} & \textbf{1.7} & \textbf{1.5} & \textbf{1.4} & \textbf{4.2} & \textbf{3.8} & \textbf{13.8} & \textbf{14.1} \\
    \bottomrule
  \end{tabular}
\end{table*}

\section{Experimental setup}


We use the same ECAPA2 model as used in Section~\ref{sec:weakly_supervised_vad} for our experimental evaluation. VAD weights and speaker embeddings are extracted using the single-step method as described in Section~\ref{sec:speaker_diarization} with a sliding window length of 2 seconds and step size of 1 second. Hysteresis thresholding is applied to determine the continuous speech segments. Afterwards, speech segments close together are merged and short segments are removed. All thresholds are determined using the development partitions of the diarization datasets. 

Spectral clustering is used to group the extracted embeddings and assign labels to the corresponding speech segments. Only speaker embeddings containing speech in the previous step are included. The affinity matrix is constructed using the cosine similarity of the speaker embeddings. We only keep the 10 highest similarities for each row in the affinity matrix for robustness. Subsequently, the eigenvectors and eigenvalues of the normalized Laplacian, derived from the pruned similarity matrix, are calculated. To estimate the number of speakers $n$ present in the input utterance, we use the eigengap criterion~\cite{spectral_clustering}. Finally, the $n$ first eigenvectors are used for K-means clustering to determine the speaker labels.

We use the evaluation partitions of the AMI, VoxConverse and DIHARD III diarization datasets to asses the performance of our proposed method. The AMI dataset consists of meeting data while VoxConverse contains multi-speaker audio collected from YouTube. DIHARD III is composed of multiple challenging domains, ranging from clinical interviews to telephone speech~\cite{dihard_iii}. For AMI, we follow the standard evaluation setup as described in~\cite{ecapa_tdnn_diarization} by using the "Full-corpus-ASR partition" and exclude the TNO meetings for both the development and test set. The audio from the microphone array is beamformed using the open-source BeamformIt toolkit~\cite{beamformit}. The diarization and VAD error rate is measured on the non-overlapping speech regions while using a forgiveness collar value of 0.25 seconds. For VoxConverse and DIHARD III, we follow the official corresponding challenge setups~\cite{voxsrc23_dia_1st, dihard_iii} by evaluating all speech regions and applying a collar value of 0.25 and 0 seconds, respectively. We note that our approach currently does not handle overlapping speech, imposing a penalty on VoxConverse and DIHARD III results. The VAD error rate is reported with false alaram (FA) and missed detection (MS) breakdown.

\section{Results}
Table~\ref{tab:vad_results} depicts the VAD error rate on all test sets of our attention-based VAD approach and other publicly available VAD models. Notably, we see that the ECAPA2 VAD system gains similar or better results in comparison to supervised models from the SpeechBrain and Silero frameworks. This indicates that the attention mechanism of a speaker embedding extractor can act as a weakly supervised VAD system with on par or better performance than comparable supervised VAD models. For reference, we also included the results of the popular PyAnnote~\cite{vad_pyannote_3.1} library, which outperforms all other models. However, it is important to notice that the PyAnnote model is trained using the in-domain development partitions of the AMI, VoxConverse and DIHARD III datasets, resulting in significant performance increases on the corresponding test sets~\cite{vad_pyannote_3.1}.


To gain an individual assessment on the diarization task of ECAPA2 speaker embeddings combined with spectral clustering, DER results using oracle VAD on the AMI and VoxConverse test sets are available in Table~\ref{tab:dia_results}, along with the best published performance of current state-of-the-art models. For a fair comparison, we only included results of non-fusion and single-scale diarization systems. We observe that our ECAPA2-based system improves the DER on average over the best models with 20.4\% and 13.7\% relative on all test sets, respectively. This illustrates that the state-of-the-art speaker verification performance of the ECAPA2 embeddings results in an accompanying increase in diarization robustness. Notably, our ECAPA2 embedding extractor achieves these results while being trained on a significantly smaller training set than all other models in Table~\ref{tab:dia_results}. We also observe that estimating the number of speaker clusters using the eigengap criterion often results in slightly better performance when compared to using the oracle number of speakers. We noted that speaker embeddings from challenging audio conditions tend to create an additional noisy cluster, preventing them from interfering with the true speaker clusters. Usage of the oracle number of speakers enforces the algorithm to assign noisy embeddings to speakers, lowering the overall robustness of the clustering results.

\section{Conclusion}
In this paper, we provided an extensive analysis of the behavior of frame-level attention weights of a speaker verification model. We established that the attention mechanism acts similarly to a weakly supervised VAD system and achieves similar or better performance as comparable supervised VAD models. Subsequently, we proposed a diarization approach which extracts the VAD weights and speaker embedding simultaneously using a single speaker verification model, providing a compelling and efficient alternative to standard cascaded speaker diarization.

\bibliographystyle{IEEEbib}
\bibliography{Odyssey2024_BibEntries}

%

\end{document}